\documentclass[
  twocolumn,
  prl,
  amsmath,
  amssymb,
  superscriptaddress,
  nofootinbib,
  floatfix
]{revtex4}
\usepackage{mathtools}
\usepackage{bm}
\usepackage{dsfont}
\usepackage{graphicx}
\usepackage{pifont}
\usepackage{textcomp}
\usepackage{gensymb}
\usepackage{accents}
\usepackage{upgreek}
\usepackage{array}
\usepackage{hhline}
\usepackage{tabularx}
\usepackage{color}
\usepackage{multirow}

\makeatletter
\def\NAT@bibsetnum#1{%
 \setlength{\topsep}{\z@}%
 \NATx@bibsetnum{#1}%
}%
\renewenvironment{thebibliography}[1]{%
 \NAT@thebibliography{#1}%
 \@clubpenalty\clubpenalty
 \let\@TBN@opr\present@bibnote
 \@FMN@list
}{%
 \@endnotesinbib
 \edef\@currentlabel{\arabic{NAT@ctr}}%
 \NAT@endthebibliography
 \global\let\auto@bib\@empty
}
\newcommand*{\supplementarystart}{%
  \close@column@grid%
  \clearpage%
  \onecolumngrid%
  \setcounter{enumiv}{0} 
  \setcounter{equation}{0} 
  \setcounter{figure}{0} 
  \setcounter{table}{0} 
  \setcounter{page}{1}
  \c@secnumdepth=4
  \renewcommand{\theequation}{s\arabic{equation}} 
  \renewcommand{\bibnumfmt}[1]{[s##1]} 
  \renewcommand{\@onlinecite}{
  \citealp} 
  \renewcommand{\cite}[1]{{[}\onlinecite{##1}{]}}
  \renewcommand{\thefigure}{s\arabic{figure}}
  \renewcommand{\thetable}{s\Roman{table}}
  \renewcommand{\thepage}{s\arabic{page}}
}
\makeatother

\newcommand{\be}{\begin{equation}}
\newcommand{\e}{\end{equation}}
\newcommand{\beml}{\begin{subequations}}
\newcommand{\eml}{\end{subequations}}
\newcommand{\beq}{\begin{eqnarray}}
\newcommand{\eq}{\end{eqnarray}}
\newcommand{\ba}{\begin{array}}
\newcommand{\ea}{\end{array}}
\newcommand{\bpm}{\begin{pmatrix}}
\newcommand{\epm}{\end{pmatrix}}
\newcommand{\bc}{\begin{cases}}
\newcommand{\ec}{\end{cases}}

\newcommand{\bb}{\boldsymbol}

\renewcommand{\log}{\mathop{\mathrm{ln}}\nolimits}

\DeclareMathOperator{\tr}{Tr}

\DeclareMathOperator{\im}{Im}
\DeclareMathOperator{\re}{Re}

\DeclareMathOperator{\li}{Li_2}

\begin{document}

\title{Asymmetric and symmetric exchange in a generalized 2D Rashba ferromagnet}

\author{I.\,A.~Ado}
\affiliation{Radboud University, Institute for Molecules and Materials, NL-6525 AJ Nijmegen, The Netherlands}

\author{A.~Qaiumzadeh}
\affiliation{Center for Quantum Spintronics, Department of Physics, Norwegian University of Science and Technology, NO-7491 Trondheim, Norway}
\affiliation{Department of Physics, Institute for Advanced Studies in Basic Sciences (IASBS), Zanjan 45137-66731, Iran}

\author{R.\,A.~Duine}
\affiliation{Center for Quantum Spintronics, Department of Physics, Norwegian University of Science and Technology, NO-7491 Trondheim, Norway}
\affiliation{Institute for Theoretical Physics and Centre for Extreme Matter and Emergent Phenomena,
Utrecht University, 3584 CE Utrecht, The Netherlands}
\affiliation{Department of Applied Physics, Eindhoven University of Technology,
P.O. Box 513, 5600 MB Eindhoven, The Netherlands}

\author{A.~Brataas}
\affiliation{Center for Quantum Spintronics, Department of Physics, Norwegian University of Science and Technology, NO-7491 Trondheim, Norway}

\author{M.~Titov}
\affiliation{Radboud University, Institute for Molecules and Materials, NL-6525 AJ Nijmegen, The Netherlands}
\affiliation{ITMO University, Saint Petersburg 197101, Russia}

\begin{abstract}
Dzyaloshinskii-Moriya interaction (DMI) is investigated in a 2D ferromagnet (FM) with spin-orbit interaction of Rashba type at finite temperatures. The FM is described in the continuum limit by an effective $s$-$d$ model with arbitrary dependence of spin-orbit coupling (SOC) and kinetic energy of itinerant electrons on the absolute value of momentum. In the limit of weak SOC, we derive a general expression for the DMI constant $D$ from a microscopic analysis of the electronic grand potential. We compare $D$ with the exchange stiffness $A$ and show that, to the leading order in small SOC strength~$\alpha_{\text{\tiny R}}$, the conventional relation $D=(4 m\alpha_{\text{\tiny R}}/\hbar)A$, in general, does not hold beyond the Bychkov-Rashba model. Moreover, in this model, both $A$ and $D$ vanish at zero temperature in the metal regime (i.\,e.,~when two spin sub-bands are partly occupied). For nonparabolic bands or nonlinear Rashba coupling, these coefficients are finite and acquire a nontrivial dependence on the chemical potential that demonstrates the possibility to control the size and chirality of magnetic textures by adjusting a gate voltage.
\end{abstract}

\maketitle
Chiral magnetic structures have attracted a great deal of interest in recent years with the observation of novel exotic magnetic phases such as skyrmion lattices~\cite{skyrmion2009}, single skyrmions~\cite{Fert-Rashba3, Hoffmann, singleskyrmions}, chiral domain walls~\cite{Thiaville, Emori, chiralDWs}, chiral magnons~\cite{chiralDWs,AFMDMI-exp,Xiao}, and helimagnets~\cite{helix}. The source of chiral symmetry breaking, required for the formation of such structures, is the asymmetric exchange interaction that is referred to as Dzyaloshinskii-Moriya interaction (DMI)~\cite{Fert-Rashba3, Dzyaloshinsky, Moriya, iDMI-theo1,iDMI-theo2,iDMI-theo3,iDMI-theo4,Lifshitz}. DMI originates from spin-orbit coupling (SOC) in magnetic systems with broken inversion symmetry, e.\,g., in noncentrosymmetric crystals or at surfaces and interfaces of thin magnetic films. The latter, effectively low-dimensional systems, which are of particular interest for applications, are in the focus of our study.

Recently, both bulk and interfacial DMI have been measured by employing Brillouin light scattering and, indirectly, using spin-polarized electron energy-loss spectroscopy~\cite{iDMI-exp0,iDMI-exp1,iDMI-exp2,iDMI-exp3,iDMI-exp4,iDMI-exp5,iDMI-exp6,iDMI-exp7,iDMI-exp8,iDMI-exp9,iDMI-exp10,iDMI-exp13}. On the other hand, for calculation of DMI in realistic materials, there exist effective computational techniques that provide decent agreement with certain experimental data~\cite{iDMI-abinitio1,Katsnelson,Ebert,iDMI-abinitio3}. A comprehensive understanding of the asymmetric exchange in generic systems requires model studies as well.

A widely used strategy for addressing DMI in systems with magnetic order is to utilize an $s$-$d$ type model approach with noninteracting itinerant electrons mediating magnetic interactions. Within this ideology, the authors of Ref.~[\onlinecite{Rashba-DMI-discrete-1}] derived formulas for the asymmetric exchange between two single magnetic ions embedded in a 1D- or 2DEG with Rashba SOC. A decade later, their result was generalized by allowing for finite uniform magnetization~\cite{Rashba-DMI-discrete-2}.

As far as smooth noncollinear magnetic structures are concerned (e.\,g., domain walls or skyrmions), it is more convenient to describe a magnet in the continuum limit by sending the lattice spacing to zero in the first place. In this paradigm, Berry phase type expressions for the asymmetric exchange have been recently derived~\cite{Mokrousov-Berry_phase} and the relation between DMI and ground-state spin currents has been pointed out~\cite{Tatara-spin_currents, Mokrousov-spin_currents}. Surprisingly, though, the only 2D ferromagnet (FM) model for which DMI has, so far, been calculated in the continuum limit refers to the system of a FM deposited on top of a topological insulator~\cite{iDMI-theory-TI1,iDMI-theory-TI2,iDMI-theory-TI3}.

In this Letter, we focus on a less exotic model that captures the effects of both Rashba SOC and the $s$-$d$ type exchange interaction between localized FM spins and 2DEG. The following Hamiltonian of one conduction electron is considered:
\be
\label{Hamiltonian}
\mathcal{H}=\xi(p)+\alpha_{\text{\tiny R}}\zeta(p)\, [\bb{p}\times\bb{\sigma}]_z  + J_{\text{sd}} S\,\bb n(\bb r,t)\cdot\bb{\sigma},
\e
where $\xi(p)$ and $\zeta(p)$ are arbitrary functions of the absolute value of momentum that parametrize free electron dispersion (kinetic energy) and momentum dependent Rashba SOC, respectively. The last term stands for the effective $s$-$d$ exchange interaction with strength~$J_{\textrm{sd}}$. We assume that the system is deep in the FM phase and the temperature is far below the corresponding Curie temperature; hence, the localized spins of the absolute value $S$ can be described by the continuous vector field $\bb n(\bb r,t)$ with the constraint $\vert\bb n\vert\equiv 1$. We also assume the dynamics of itinerant electrons to be much faster than that of FM spins and treat the field $\bb n$ as time independent. The notation $\bb\sigma$ refers to a vector of Pauli matrices. 

The model of Eq.~(\ref{Hamiltonian}) describes a generic FM layer coupled to 2DEG with spin-orbit interaction of Rashba type. One possible realization of such a system is a LaAlO$_3$/SrTiO$_3$ interface~\cite{LaAlO3...2010}. The model might also be used to describe a SrRuO$_3$/SrIrO$_3$ interface, which has recently gained considerable attention in the context of the so-called topological Hall effect -- the phenomenon intrinsically linked to DMI~\cite{SrRuO3...2016,SrRuO3...2018}. 

In the continuum limit, DMI (or the antisymmetric exchange) is recognized as a contribution $\Omega_D[\bb n]$ to the micromagnetic free energy density that is linear with respect to the first spatial derivatives of the vector field~$\bb n$. The symmetric exchange, on the other hand, is associated with a contribution $\Omega_A[\bb n]$ that is quadratic with respect to the first spatial derivatives of $\bb n$. The ratio between the two contributions plays a key role in formation of chiral magnetic structures, affecting their stability and size. Relation between $\Omega_D[\bb n]$ and $\Omega_A[\bb n]$ for the model of Eq.~(\ref{Hamiltonian}) is interesting for one more, historical, reason. Standard symmetry analysis shows~\cite{Lifshitz} that in an isotropic 2D FM system, one has
\be
\label{symmetric_energy_general}
\Omega_A[\bb n]=A\left[(\nabla_x\bb n)^2+(\nabla_y\bb n)^2\right],
\e
where $A$ is the exchange stiffness. For a particular choice, $\xi(p)=p^2/2m$ and $\zeta(p)\equiv 1$ in Eq.~(\ref{Hamiltonian}), which is referred to below as 
the Bychkov-Rashba model~\cite{Rashba}, the authors of Ref.~[\onlinecite{Stiles}] argued that, in the limit of weak SOC, the form of Eq.~(\ref{symmetric_energy_general}) necessarily leads to
\be
\label{asymmetric_energy_general}
\Omega_D[\bb n]=D\,\bb n \cdot [[\bb e_z\times\bb\nabla]\times\bb n]
\e
and, moreover, to $D=(4m\alpha_{\text{\tiny R}}/\hbar)A$. Unfortunately, the actual calculation of $\Omega_D[\bb n]$ has been performed neither in Ref.~[\onlinecite{Stiles}] nor, to the best of our knowledge, anywhere else even for the particular case of the Bychkov-Rashba model.

Below, we undertake an accurate microscopic treatment of the model of Eq.~(\ref{Hamiltonian}) in the leading order with respect to small $\alpha_{\text{\tiny R}}$ and, under rather general assumptions on $\xi(p)$ and $\zeta(p)$ \cite{supp}, directly derive Eqs.~(\ref{symmetric_energy_general}) and~(\ref{asymmetric_energy_general}). Furthermore, we report that the exchange stiffness $A$ and the DMI constant $D$ are given by remarkably concise expressions, namely,
\begin{gather}
\label{A}
A=\frac{\Delta_{\text{sd}}}{32\pi}
\frac{\partial}{\partial \Delta_{\text{sd}}}
\left[
\int_0^{\infty}{dp\,
\frac{p\,[\xi'(p)]^2}{\Delta_{\text{sd}}}(f_--f_+)}
\right],
\\
\label{D}
D=\frac{\alpha_{\text{\tiny R}}\Delta_{\text{sd}}}{8\pi\hbar}
\frac{\partial}{\partial \Delta_{\text{sd}}}
\left[
\int_0^{\infty}{dp\,
\frac{p^2\,\zeta(p)\xi'(p)}{\Delta_{\text{sd}}}(f_--f_+)}
\right],
\end{gather}
where $\Delta_{\text{sd}}=\vert J_{\text{sd}}\vert S$ is half of the exchange splitting, $\xi'(p)=\partial\xi/\partial p$, and $f_{\pm}=f(\xi(p)\pm\Delta_{\text{sd}})$ are expressed via the Fermi-Dirac distribution
\be
f(\varepsilon)=\left(1+\exp{\left[(\varepsilon-\mu)/T\right]}\right)^{-1}
\e
with the chemical potential $\mu$ and temperature $T$.

We would like to draw the reader's attention to the fact that the result of Eq.~(\ref{A}) is well-known, though, in a different form (see, e.\,g., Eq.~(70) in Ref.~[\onlinecite{Katsnelson's_review}]). It is, however, useful to cast $A$ in the form of Eq.~(\ref{A}) in order to compare the symmetric and asymmetric exchange for several particular choices of $\xi(p)$ and $\zeta(p)$ as we do later in the text.

We have checked that the DMI constant of Eq.~(\ref{D}) can also be obtained either by evaluation of ground-state spin currents~\cite{Tatara-spin_currents, Mokrousov-spin_currents} or by using the formalism of Ref.~[\onlinecite{Mokrousov-Berry_phase}]. We have also checked that one may restore both Eqs.~(\ref{A}) and (\ref{D}) by calculation of spin density of conduction electrons $\bb s$~\cite{comment3} followed by an integration of the relation $\bb n\times(\delta\Omega/\delta \bb n)=J_{\text{sd}} S\,\bb n\times\bb s$, as it was done in Ref.~[\onlinecite{iDMI-theory-TI1}] for DMI in the Dirac model. It must also be possible to compute $A$ and $D$ from an effective action~\cite{iDMI-theory-TI3,RashbaAFM-DMI}.

Nevertheless, we believe that the most natural and straightforward way to derive Eqs.~(\ref{A}) and (\ref{D}) is to extract $\Omega_A[\bb n]$ and $\Omega_D[\bb n]$ from the electronic grand potential density $\Omega$. In this approach, there is no need to assume \textit{a priori} the symmetry form of the final result as it is often done in the literature. Using the standard formulation of statistical physics, we express the grand potential density at $\bb r = \bb r_0$ as
\be
\label{Omega_general}
\Omega=-\frac{T}{2\pi i}\tr{\int\limits{d \varepsilon\,g(\varepsilon)\left[\mathcal G^A(\bb r_0, \bb r_0)-\mathcal G^R(\bb r_0, \bb r_0)\right]}},
\e
where $\mathcal{G}^{A(R)}=(\varepsilon\mp i 0-\mathcal H)^{-1}$ is the advanced (retarded) Green's function for the model of Eq.~(\ref{Hamiltonian}), $\tr$ stands for the matrix trace, and the notation
\be
g(\varepsilon)=\log{\left(1+\exp{\left[(\mu-\varepsilon)/T\right]}\right)}
\e
is employed.

Now, let us show how Eq.~(\ref{Omega_general}) can be used to obtain the DMI contribution to micromagnetic free energy density. First, one should Taylor expand 
$\bb n(\bb r)$ around $\bb n(\bb r_0)$ and use the result to generate the Dyson series
\begin{multline}
\label{Dyson}
\mathcal G(\bb r_0,\bb r_0)=G(\bb r_0-\bb r_0)+
J_{\text{sd}} S\int d\bb r'\,
G(\bb r_0-\bb r')\\\times
\left[
\sum\limits_{\beta\gamma}(\bb r'-\bb r_0)_\beta\nabla_\beta\, n_\gamma(\bb r_0)\,\sigma_\gamma
\right]
G(\bb r'-\bb r_0),
\end{multline}
where $G$ is the Green's function of a homogeneous system with fixed $\bb n(\bb r)\equiv\bb n(\bb r_0)$. In Eq.~(\ref{Dyson}), we have disregarded all the gradients of $\bb n$ but the first, which is only accounted for in the linear order. The second term in Eq.~(\ref{Dyson}) is precisely the one that determines the asymmetric exchange. Substituting it into Eq.~(\ref{Omega_general}), we switch to momentum representation and symmetrize the result to obtain the general formula
\be
\label{DMI_general_0}
\Omega_D[\bb n]=
\sum_{\beta\gamma}{\Omega^{\text{DMI}}_{\beta\gamma}\,\nabla_\beta\, n_\gamma},
\e
with the DMI tensor defined as
\begin{multline}
\label{DMI_general}
\phantom{\Biggr|}
\Omega^{\text{DMI}}_{\beta\gamma}=T\frac{J_{\text{sd}} S}{2\pi\hbar}
\re
\int d \varepsilon\,
g(\varepsilon)
\int
\frac{d^2 p}{(2\pi)^2}\\
\times\tr{\Bigl(
G^{R}\sigma_\gamma\,G^{R}\,v_\beta\,G^{R}-G^{R}\,v_\beta\,G^{R}\sigma_\gamma\,G^{R}
\Bigr)},
\end{multline}
where $\bb v=\partial \mathcal H/\partial \bb p$ is the velocity operator. Note that we have dropped the argument of $\bb n(\bb r_0)$ in Eq.~(\ref{DMI_general_0}) and further below.

Evaluation of Eq.~(\ref{DMI_general}) for the present model 
is performed with the help of the momentum-dependent Green's function
\be
\label{green's_functions}
G^{R(A)}=\frac{\varepsilon-\xi(p)+\alpha_{\text{\tiny R}}\zeta(p)\, [\bb{p}\times\bb{\sigma}]_z  + J_{\text{sd}}S\,\bb n\cdot\bb{\sigma}}{(\varepsilon-\varepsilon^+(\bb p)\pm i0)(\varepsilon-\varepsilon^-(\bb p)\pm i0)},
\e
where we introduce the spectral branches $\varepsilon^{\pm}(\bb p)=\xi(p)\pm\sqrt{(J_{\text{sd}} S)^2+[\alpha_{\text{\tiny R}}p\,\zeta(p)]^2-2\alpha_{\text{\tiny R}}J_{\text{sd}} S \,p\,\zeta(p)\sin{\theta}\sin{\phi}}$, the angle $\theta$ stands for the polar angle of $\bb n$ with respect to the $z$ axis, while $\phi$ is the angle between the momentum $\bb{p}$ and the in-plane projection of the vector $\bb{n}$. We substitute Eq.~(\ref{green's_functions}) into  Eq.~(\ref{DMI_general}), calculate the matrix trace, expand the integrands to the linear order in $\alpha_{\text{\tiny R}}$, and straightforwardly integrate over $\phi$. This results in the following form of the DMI tensor:
\be
\label{calculated_Omega_D}
\Omega^{\text{DMI}}_{\beta\gamma}=D\,
\sum\limits_{i j}{n_i\epsilon_{i j\gamma}\epsilon_{j z \beta}},
\e
where $\epsilon_{q_1 q_2 q_3}$ denotes the three-dimensional Levi-Civita symbol, while
\begin{multline}
\label{almost_D}
D=\frac{\alpha_{\text{\tiny R}}\Delta_{\text{sd}}^2}{2\pi^2\hbar}T
\int_0^{\infty}p\,dp\int_{-\infty}^{\infty}d \varepsilon\,
g(\varepsilon)\\
\times\im{\left(
\frac{2\zeta(p)+p\,\zeta'(p)}{[\varepsilon-\varepsilon^{+}_0(\bb p)+i 0]^2[\varepsilon-\varepsilon^{-}_0(\bb p)+i 0]^2}
\right)},
\end{multline}
where $\zeta'(p)=\partial\zeta/\partial p$ and $\varepsilon^{\pm}_0(\bb p)=\xi(p)\pm \Delta_{\text{sd}}$. From Eqs.~(\ref{calculated_Omega_D}) and (\ref{almost_D}), it is already evident that, up to the linear order in $\alpha_{\text{\tiny R}}$, the asymmetric exchange does, indeed, have the form of Eq.~(\ref{asymmetric_energy_general}) with the DMI constant $D$ which is totally independent of the direction of magnetization.

Integration over $\varepsilon$ in Eq.~(\ref{almost_D}) leads to
\begin{multline}
D=\frac{\alpha_{\text{\tiny R}}\Delta_{\text{sd}}}{8\pi\hbar}T
\int_{0}^{\infty}{d p\,\frac{g_-' - g_+'}{\Delta_{\text{sd}}}\frac{\partial\left[p^2\zeta(p)\right]}{\partial p}}
\\
-
\frac{\alpha_{\text{\tiny R}}\Delta_{\text{sd}}}{8\pi\hbar}T
\int_{0}^{\infty}{d p\,\frac{g_- - g_+}{\Delta_{\text{sd}}^2}\frac{\partial\left[p^2\zeta(p)\right]}{\partial p}},
\end{multline}
where $g_{\pm}'=\partial g_{\pm}/\partial \Delta_{\text{sd}}$ and $g_{\pm}=g(\xi(p)\pm\Delta_{\text{sd}})$~\cite{comment3.9}. Eventually, the above two integrals are combined to form a full derivative with respect to $\Delta_{\text{sd}}$. Partial integration concludes the derivation of the DMI constant $D$ of Eq.~(\ref{D}) once the identity $\partial g(\varepsilon)/\partial \varepsilon=-f(\varepsilon)/T$ is used.

The symmetric exchange can be treated similarly. In~order to derive Eqs.~(\ref{symmetric_energy_general}) and (\ref{A}), one should take $\alpha_{\text{\tiny R}}=0$ and extract all terms proportional to $\nabla_\beta n_\gamma\nabla_{\beta'} n_{\gamma'}$ and $\nabla_\beta\nabla_{\beta'} n_\gamma$ in Eq.~(\ref{Omega_general}). We relegate the details of the calculation to the Supplementary Material~\cite{supp}.

\begin{figure}[t!]
\includegraphics[width=0.95\columnwidth]{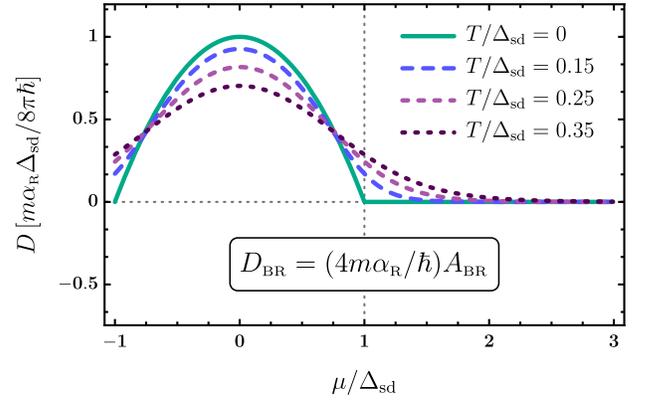}
\caption{Dzyaloshinskii-Moriya interaction constant $D$ in the Bychkov-Rashba model as a function of the chemical potential $\mu$ at different temperatures $T$. Both $\mu$ and $T$ are normalized by half of the exchange splitting $\Delta_{\text{sd}}=\vert J_{\text{sd}}\vert S$.}
\label{fig::D-parabolic}
\end{figure}

\begin{figure}[t!]
\includegraphics[width=0.95\columnwidth]{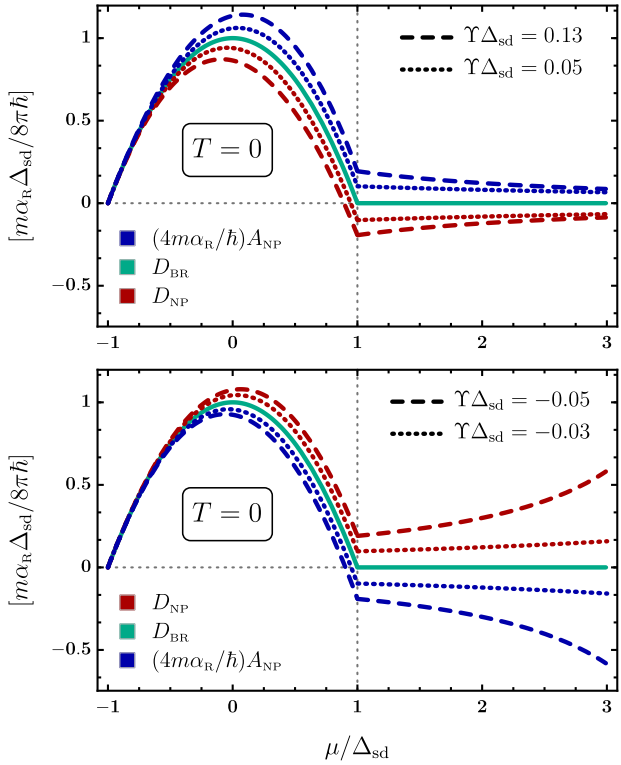}
\caption{Dzyaloshinskii-Moriya interaction constant $D$ and ``normalized'' exchange stiffness $(4 m\alpha_{\text{\tiny R}}/\hbar)A$ as functions of the chemical potential $\mu$ at zero temperature for different values of nonparabolicity coefficient $\Upsilon$. Both $\mu$ and $1/\Upsilon$ are normalized by half of the exchange splitting $\Delta_{\text{sd}}=\vert J_{\text{sd}}\vert S$.}
\label{fig::DvsA}
\end{figure}

\begin{figure}[b!]
\includegraphics[width=0.95\columnwidth]{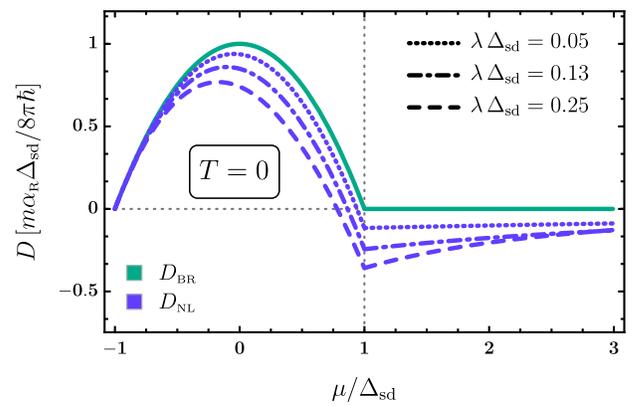}
\caption{Dzyaloshinskii-Moriya interaction constant $D$ as a function of the chemical potential $\mu$ at zero temperature for different values of nonlinearity coefficient $\lambda$. Both $\mu$ and $1/\lambda$ are normalized by half of the exchange splitting $\Delta_{\text{sd}}=\vert J_{\text{sd}}\vert S$.}
\label{fig::D_nonlinear_alpha}
\end{figure}

In the rest of the Letter, we apply the general expressions of Eqs.~(\ref{A}) and (\ref{D}) to three particular cases. All further analytical results are presented in Table~\ref{table}, and the corresponding plots are given in Figs.~\ref{fig::D-parabolic},~\ref{fig::DvsA}, and \ref{fig::D_nonlinear_alpha}.

To begin with, we return to the Bychkov-Rashba (BR) model characterized by $\xi(p)=p^2/2m$ and $\zeta(p)\equiv 1$. As can be immediately seen from Eqs.~(\ref{A}) and (\ref{D}), the relation $D_{\text{\tiny{BR}}}=(4 m\alpha_{\text{\tiny R}}/\hbar)A_{\text{\tiny{BR}}}$, indeed, holds, and the prediction of Ref.~[\onlinecite{Stiles}] is validated. Furthermore, in the limit of zero temperature, one finds from Eq.~(\ref{D}) that
\be
D_{\text{\tiny{BR}}}=
\frac{\Delta_{\text{sd}} m\alpha_{\text{\tiny{R}}}}{8\pi\hbar}
\begin{cases}
1-\left(\mu/\Delta_{\text{sd}}\right)^2,
&\vert\mu\vert<\Delta_{\text{sd}}\\
0, &\phantom{\vert}\mu\phantom{\vert}>\Delta_{\text{sd}}
\end{cases}
.
\e
Thus, if SOC is weak, both $A$ and $D$ are finite in the Bychkov-Rashba model at $T=0$ only in the half-metal regime $\vert\mu\vert<\Delta_{\text{sd}}$.

In fact, DMI in this model vanishes identically in the metal regime $\mu>\Delta_{\text{sd}}$ irrespective of the SOC strength. At larger~$\alpha_{\text{\tiny{R}}}$, the asymmetric exchange ceases to have the simple symmetry of Eq.~(\ref{asymmetric_energy_general}) in the form of Lifshitz invariants. However, contributions from the two Fermi surfaces still cancel each other within each component of the DMI tensor $\Omega^{\text{DMI}}$, no matter what the SOC strength is~\cite{comment4,comment5,Yudin}. A nonperturbative in SOC study of DMI in the model of Eq.~(\ref{Hamiltonian}) will be presented elsewhere.

Next, it is instructive to see how the deviations from parabolic dispersion, a common property of, e.\,g., narrow gap semiconductors and quantum wells~\cite{{nonparabolicity-I},{nonparabolicity-II},{nonparabolicity-III}}, affect $A$ and $D$ and the relation between them. To model nonparabolicity (NP) we use $\xi(p)=(p^2/2m)(1+\Upsilon\, p^2/2m)$ and $\zeta(p)\equiv 1$ with the parameter $\Upsilon$ quantifying the deviation from the parabolic band. We shall assume that $\xi(p)$ is an increasing function even for negative values of~$\Upsilon$; i.\,e., our choice of $\xi(p)$ is understood as an approximation at small values of $p$. Temperature is set to zero.

We find, in this case, that the DMI constant and the exchange stiffness remain finite for all values of $\mu$. Moreover, the NP corrections to $D$ and $(4 m\alpha_{\text{\tiny R}}/\hbar)A$ are of different signs, but have equal magnitudes,
\be
D_{\text{\tiny{NP}}}-D_{\text{\tiny{BR}}}=-(4 m\alpha_{\text{\tiny R}}/\hbar)(A_{\text{\tiny{NP}}}-A_{\text{\tiny{BR}}}),
\e
independently of the sign of $\Upsilon$ (see Fig.~\ref{fig::DvsA} and Table~\ref{table}). This leads, in the metal regime, to a particularly unexpected relation
\be
D_{\text{\tiny{NP}}}=-(4 m\alpha_{\text{\tiny R}}/\hbar)A_{\text{\tiny{NP}}},\quad \mu>\Delta_{\text{sd}}
\e
(cf. the relation $D_{\text{\tiny{BR}}}=(4 m\alpha_{\text{\tiny R}}/\hbar)A_{\text{\tiny{BR}}}$ for the Bychkov-Rashba model).

For $\Upsilon<0$, the exchange stiffness becomes negative in the metal regime, which may eventually make the FM phase unstable. Of course, within our study, we do not consider direct contributions to magnetic exchange that may remain sufficiently large to be overcome by negative~$A_{\text{\tiny{NP}}}$. Nevertheless, the reduction of the direct exchange in nonparabolic FM layers may have a serious impact on the size of noncollinear magnetic textures. In a particular case of a single skyrmion, a simple estimate of its size is $\propto A/D$~\cite{size_of_skyrmion}. We note that, for $\Upsilon<0$, the DMI constant is enhanced; hence, the deviations from parabolicity may reduce the size of magnetic skyrmions leading to miniaturization of skyrmion-based technology. In general, nontrivial dependence of $A$ and $D$ on the chemical potential shown in Fig.~\ref{fig::DvsA} clearly demonstrates the possibility to control the size of skyrmions by means of a gate voltage.

Finally, motivated by theoretical~\cite{nonlinear_RSOC_theory}, computational~\cite{nonlinear_RSOC_DFT}, and experimental~\cite{nonlinear_RSOC_exp} demonstrations of generally nonlinear (NL) dependence of Rashba SOC on momentum, we model the effect of the latter on the asymmetric exchange. Since Rashba spin splitting is usually reported~\cite{nonlinear_RSOC_theory,nonlinear_RSOC_DFT,nonlinear_RSOC_exp} to either saturate or decrease with increasing $p$, we use $\zeta(p)=1/\left(1+\lambda\, p^2/2m\right)$ with positive parameter $\lambda$ and $\xi(p)=p^2/2m$. At zero temperature, we then find a finite DMI constant $D_{\text{\tiny{NL}}}$ for any value of the chemical potential (see Fig.~\ref{fig::D_nonlinear_alpha} and Table~\ref{table}). Moreover, $D_{\text{\tiny{NL}}}$ exhibits a sign change around $\mu=\Delta_{\textrm{sd}}$. This demonstrates that a gate voltage can also be used to manipulate chirality of magnetic order in 2D FM.

\begingroup
\begin{table*}
\begin{center}
{\renewcommand{\arraystretch}{1.2}
\tabcolsep=2.4pt
\begin{tabular}{c|c|c|c|cc|c}
\hline
\multicolumn{3}{c}{\multirow{2}{*}{Cases}}
	& \multicolumn{1}{c}{}
		& \multicolumn{3}{c}{$D$ and $\left(4 m\alpha_{\text{\tiny R}}/\hbar\right)A$ in 
		units $m\alpha_{\text{\tiny{R}}}\Delta_{\text{sd}}/8\pi\hbar$}\\
\cline{5-7}
\multicolumn{3}{c}{}&\multicolumn{1}{c}{}
			& \multicolumn{2}{c|}{$\vert\mu\vert<\Delta_{\text{sd}}$}
				& $\mu>\Delta_{\text{sd}}$\\
\hline
\multicolumn{3}{c|}{\multirow{2}{*}{$\xi(p)=p^2/2m$, $\zeta(p)\equiv 1$\hspace{29pt}}} 
	& $D$
			& \multicolumn{3}{c}{\multirow{2}{*}
			{$1-\left(\mu/\Delta_{\text{sd}}\right)^2-
			(\pi^2/3)\left(T/\Delta_{\text{sd}}\right)^2+
			S\left(\mu/\Delta_{\text{sd}},
			T/\Delta_{\text{sd}}\right)$}}\\
\cline{4-4}
\multicolumn{3}{c|}{}
	& $\left(4 m\alpha_{\text{\tiny R}}/\hbar\right)A$ \\
\hhline{=======}
\multicolumn{2}{c|}{\multirow{2}{*}{$\xi(p)=p^2/2m$, $\zeta(p)\equiv 1$}} &
  \multirow{6}{*}{\vspace{1ex}$T=0$}
	& $D$ 
			& \multicolumn{2}{c|}{\multirow{2}{*}			
			{$1-\left(\mu/\Delta_{\text{sd}}\right)^2$}}
				& \multirow{2}{*}{0}\\
\cline{4-4}
\multicolumn{2}{c|}{}&
	& $\left(4 m\alpha_{\text{\tiny R}}/\hbar\right)A$
			&
				&\\
\hhline{==|~|====}
\multicolumn{2}{c|}{\multirow{2}{*}{$\xi(p)=(p^2/2m)\left(1+\Upsilon\, p^2/2m\right)$, $\zeta(p)\equiv 1$}} &
	& $D$
			& \multicolumn{2}{c|}{$\phantom{w}1-\left(\mu/\Delta_{\text{sd}}\right)^2 -
			u\left(\mu/\Delta_{\text{sd}},
			\Upsilon\Delta_{\text{sd}}\right)\phantom{w}$}
				&$-U\left(\mu/\Delta_{\text{sd}},
				\Upsilon\Delta_{\text{sd}}\right)$\\
\cline{4-7}
\multicolumn{2}{c|}{}&
	& $\left(4 m\alpha_{\text{\tiny R}}/\hbar\right)A$
			& \multicolumn{2}{c|}{$\phantom{w}1-\left(\mu/\Delta_{\text{sd}}\right)^2 +
			u\left(\mu/\Delta_{\text{sd}},
			\Upsilon\Delta_{\text{sd}}\right)\phantom{w}$}
				&$\phantom{-}U\left(\mu/\Delta_{\text{sd}},
				\Upsilon\Delta_{\text{sd}}\right)$\\
\hhline{==|~|====}
\multicolumn{2}{c|}{$\xi(p)=p^2/2m$, $\zeta(p)=1/\left(1+\lambda\, p^2/2m\right)$} &
	& $D$
			& \multicolumn{2}{c|}{$\phantom{w}1-\left(\mu/\Delta_{\text{sd}}\right)^2 -
			w\left(\mu/\Delta_{\text{sd}},
			\lambda\Delta_{\text{sd}}\right)\phantom{w}$}
				&$-W\left(\mu/\Delta_{\text{sd}},
				\lambda\Delta_{\text{sd}}\right)$\\
\hline
\multirow{3}{*}{Notations}&
	\multicolumn{4}{l}
	{$s\left(a,b\right)=2b\left\{\log{\left(1+\exp{\left[-\left(a+1\right)/b\,
	\right]}\right)}-b\li{\left(-\exp{\left[-\left(a+1\right)/b\,
	\right]}\right)}\right\}$,}
		& \multicolumn{2}{l}{$\phantom{hu}S\left(a,b\right)=s\left(a,b\right)+s\left(-
		a,b\right)$}\\
	&\multicolumn{4}{l}{$u\left(a,b\right)=(q-1)^2\left\{24b-(q-1)(3q+1)\right\}/48 b^2$,
	\hspace{10pt} $q=\sqrt{1+4(a+1)b}$,}
		& \multicolumn{2}{l}{$\phantom{hu}U\left(a,b\right)=u\left(a,b\right)-u\left(-a,-
		b\right)$}\\
	&\multicolumn{4}{l}{$w\left(a,b\right)=\left\{\left[2b(r-1)-r(r-3)\right]
	(r-1)-2r\log{r}\right\}/r b^2$,\hspace{10pt} $r=1+(a+1)b$,}
		& \multicolumn{2}{l}{$\phantom{hu}W\left(a,b\right)=w\left(a,b\right)-w\left(-a,-
		b\right)$}\\
\hline
\end{tabular}
}
\end{center}
\caption{\label{table}Analytical results for $D$ and $\left(4 m\alpha_{\text{\tiny R}}/\hbar\right)A$ for particular choices of $\xi(p)$ and $\zeta(p)$. Results that correspond to the Bychkov-Rashba model are shown both with full temperature dependence (upper row) and, for clarity, at zero temperature (second row). Notation $\li{(\cdot)}$ stands for the dilogarithm, which is the polylogarithm of the order $2$. Sign of $\Upsilon$ can be taken arbitrary, whereas $\lambda$ is assumed positive. For $\Upsilon<0$ the expressions shown are valid only as long as $\mu<(4\vert\Upsilon\vert)^{-1}-\Delta_{\text{sd}}$. Here signs of $\Upsilon$, $u$, and $U$ coincide, while $w$ and $W$ are positive.}
\end{table*}
\endgroup

Tuning of DMI has, so far, been realized by different approaches to interface engineering~\cite{tunability-I,tunability-II}. The ambition to manipulate the stability parameter, size, and density of skyrmions was very recently achieved as well, by means of similar methods~\cite{tunability-III}. Based on our findings, we argue that a gate voltage variation may add yet another important and flexible tool for controlling chiral magnetic domains, paving the way towards novel material design.

To conclude, we considered the asymmetric exchange in generalized 2D Rashba FM. In the weak SOC limit, we established the full form of the corresponding contribution to micromagnetic free energy density and derived a general formula for the DMI constant. We showed that, to the leading order in small $\alpha_{\text{\tiny R}}$ in the Bychkov-Rashba model, a linear relation between the exchange stiffness $A$ and the DMI constant $D$, indeed, holds, while at zero temperature, both vanish once the two spin sub-bands are partly occupied. At the same time, deviations from the Bychkov-Rashba model prevent this cancellation. There is no general linear dependence between $A$ and $D$. In particular, the relation $D=(4m\alpha_{\textrm{R}}/\hbar)A$ for the Bychkov-Rashba model is replaced at zero temperature by the relation $D=-(4 m\alpha_{\textrm{R}}/\hbar)A$ in the metal regime of the same model if nonparabolicity of the kinetic term is taken into account. For nonparabolic bands or nonlinear Rashba coupling, both $A$ and $D$ acquire a nontrivial dependence on the chemical potential that demonstrates the possibility of controlling the size and chirality of magnetic textures by adjusting a gate voltage.

\begin{acknowledgments}
We are grateful to M.\,I.~Katsnelson, P.\,M.~Ostrovsky, S.~Brener, O.~Gomonay, and K.-W.~Kim for helpful discussions. This research was supported by the European Research Council via Advanced Grant No. 669442 ``Insulatronics'', the Research Council of Norway through its Centres of Excellence funding scheme, Project No. 262633, ``QuSpin'', and by the JTC-FLAGERA Project GRANSPORT. M.T. acknowledges support from the Russian Science Foundation under Project No. 17-12-01359. R.D. is part of the D-ITP consortium, a program of the Netherlands Organisation for Scientific Research (NWO) that is funded by the Dutch Ministry of Education, Culture and Science (OCW).
\end{acknowledgments}

\supplementarystart

\centerline{\bfseries\large ONLINE SUPPLEMENTARY MATERIAL}
\vspace{6pt}
\centerline{\bfseries\large Asymmetric and symmetric exchange in a generalized 2D Rashba ferromagnet}
\vspace{6pt}
\centerline{I.\,A.~Ado, A.~Qaiumzadeh, R.\,A.~Duine, A.~Brataas, and M.~Titov}
\begin{quote}
In this Supplementary Material, we formulate the assumptions on
$\xi(p)$ and $\zeta(p)$ and also derive Eqs.~(\ref{symmetric_energy_general}) and (\ref{A}) of the main text of the Letter.
\end{quote}

\subsection{Assumptions on $\xi(p)$ and $\zeta(p)$}
The result of Eq.~(\ref{D}) assumes that the derivative $\partial D/\partial \alpha_{\text{\tiny R}}$ at $\alpha_{\text{\tiny R}}=0$ does exist. The latter is not the case, e.\,g., for the model of Dirac fermions, where $D\propto 1/\alpha_{\text{\tiny R}}$~\cite{iDMI-theory-TI1,iDMI-theory-TI2,iDMI-theory-TI3}. Thus, the necessary condition for the validity of Eq.~(\ref{D}) is $\xi(p)\not\equiv 0$. In order to establish the sufficient conditions, one should investigate the convergence of the integrals that define $\partial D/\partial \alpha_{\text{\tiny R}}$. Given $\xi(p)$ and $\zeta(p)$ have no singularities at finite values of $p$, it would be a study of convergence of the corresponding integrals at $p=\infty$. Uniform convergence is guaranteed, for instance, if distribution functions $f(\varepsilon^{\pm}(\bb p))$ decay at infinity well enough. This will be the case if at large $p$ function $\xi(p)$ is positive, unbounded, and grows faster than $\vert p\,\zeta(p)\vert$.

The result of Eq.~(\ref{A}) provides the value of the exchange stiffness in the absence of SOC, hence it depends on $\xi(\cdot)$ only. If $\xi(p)$ has no singularities at finite values of $p$, and it is positive and unbounded at large $p$, Eq.~(\ref{A}) is valid.

\subsection{Derivation of Eqs.~(\ref{symmetric_energy_general}) and (\ref{A}) of the main text of the Letter}
In order to compute the symmetric exchange contribution to micromagnetic free energy density, one has to extract all terms proportional to $\nabla_\beta n_\gamma\nabla_{\beta'} n_{\gamma'}$ and $\nabla_\beta\nabla_{\beta'} n_\gamma$ in the electronic grand potential, Eq.~(\ref{Omega_general}). To do that, we extend the Dyson series of Eq.~(\ref{Dyson}) as
\begin{multline}
\label{Dyson_hardcore}
\mathcal G(\bb r_0,\bb r_0)=G(\bb r_0-\bb r_0)+
J_{\text{sd}} S\int d\bb r'\,
G(\bb r_0-\bb r')
\left[
\sum\limits_{\beta\gamma}(\bb r'-\bb r_0)_\beta\nabla_\beta n_\gamma(\bb r_0)\,\sigma_\gamma
\right]
G(\bb r'-\bb r_0)
\\
+
(J_{\text{sd}} S)^2\int d\bb r'd\bb r''\,
G(\bb r_0-\bb r')
\left[
\sum\limits_{\beta\gamma}(\bb r'-\bb r_0)_\beta\nabla_\beta n_\gamma(\bb r_0)\,\sigma_\gamma
\right]
G(\bb r'-\bb r'')
\left[
\sum\limits_{\beta'\gamma'}(\bb r''-\bb r_0)_{\beta'}\nabla_{\beta'} n_{\gamma'}(\bb r_0)\,\sigma_{\gamma'}
\right]
G(\bb r''-\bb r_0)
\\
+
\frac{J_{\text{sd}} S}{2}\int d\bb r'\,
G(\bb r_0-\bb r')
\left[
\sum\limits_{\beta\beta'\gamma}(\bb r'-\bb r_0)_\beta(\bb r'-\bb r_0)_{\beta'}\nabla_\beta\nabla_{\beta'} n_\gamma(\bb r_0)
\right]
G(\bb r'-\bb r_0),
\end{multline}
where the first line has been already analysed in the main text, the second line is a second order correction to the Green's function due to the first spatial derivatives of $\bb n$, while the third line is a first order correction due to the second spatial derivatives of $\bb n$. We substitute the latter two into Eq.~(\ref{Omega_general}), switch to momentum representation, and symmetrize the outcome, arriving at
\be
\label{Exc_general_0}
\Omega_A[\bb n]=
\sum_{\beta\beta'\gamma\gamma'}{\Omega^{\text{exc-I}}_{\beta\beta'\gamma\gamma'}\nabla_\beta\, n_\gamma\nabla_{\beta'}\, n_{\gamma'}}
+
\sum_{\beta\beta'\gamma}{\Omega^{\text{exc-II}}_{\beta\beta'\gamma}\,\nabla_\beta\nabla_{\beta'}\, n_\gamma},
\e
where the tensors are defined as
\be
\label{Exc_general_1}
\Omega^{\text{exc-I}}_{\beta\beta'\gamma\gamma'}=
T\frac{(J_{\text{sd}} S)^2}{2\pi}
\im{
\int{d \varepsilon\,
g(\varepsilon)
\int{
\frac{d^2 p}{(2\pi)^2}
\tr{
\Bigl(
G^{R}\,v_\beta\,G^{R}\sigma_\gamma\,G^{R}\sigma_{\gamma'}\,G^{R}\,v_{\beta'}\,G^{R}+
G^{R}\,v_{\beta'}\,G^{R}\sigma_{\gamma'}\,G^{R}\sigma_\gamma\,G^{R}\,v_\beta\,G^{R}
\Bigr)
}
}
}
}
\e
and
\be
\label{Exc_general_2}
\Omega^{\text{exc-II}}_{\beta\beta'\gamma}=
-T\frac{J_{\text{sd}} S}{4\pi}
\im{
\int{d \varepsilon\,
g(\varepsilon)
\int{
\frac{d^2 p}{(2\pi)^2}
\tr{\left(
\frac{\partial^2 G^{R}}{\partial p_\beta\partial p_{\beta'}}\sigma_\gamma\,G^{R}+
G^{R}\sigma_\gamma\,\frac{\partial^2 G^{R}}{\partial p_\beta\partial p_{\beta'}}
\right)}
}
}
}.
\e
The notation of the argument of $\bb n(\bb r_0)$ is dropped in Eq.~(\ref{Exc_general_0}) and further below.

The Green's functions entering Eqs.~(\ref{Exc_general_1}) and (\ref{Exc_general_2}) are taken in the momentum representation of Eq.~(\ref{green's_functions}) of the main text, but with $\alpha_{\text{\tiny{R}}}=0$. Taking a matrix trace calculation and performing an integration over the angle, we obtain
\begin{gather}
\label{calculated_Omega_A_1}
\Omega^{\text{exc-I}}_{\beta\beta'\gamma\gamma'}=
A_1\,\delta_{\beta\beta'}\delta_{\gamma\gamma'}+
W\,\delta_{\beta\beta'}n_{\gamma}n_{\gamma'},
\\
\label{calculated_Omega_A_2}
\Omega^{\text{exc-II}}_{\beta\beta'\gamma}=
A_2\,\delta_{\beta\beta'}n_{\gamma},
\end{gather}
where $\delta_{q_1 q_2}$ is Kronecker delta, while
\begin{gather}
\label{almost_A_1}
A_1=\frac{\Delta_{\text{sd}}^2}{2\pi^2}T
\int\limits_0^{\infty}p\,dp\int\limits_{-\infty}^{\infty}d\varepsilon\,
g(\varepsilon)\,
\im{\left(
\frac{\left[\xi'(p)\right]^2\left[3\Delta_{\text{sd}}^2+(\varepsilon-\xi(p))^2\right] \left[\varepsilon-\xi(p)\right]}{[\varepsilon+i 0-\varepsilon^{+}_0(\bb p)]^4[\varepsilon+i 0-\varepsilon^{-}_0(\bb p)]^4}
\right)},
\\
\label{almost_A_2}
A_2=-\frac{\Delta_{\text{sd}}^2}{\pi^2}T
\int\limits_0^{\infty}p\,dp\int\limits_{-\infty}^{\infty}d\varepsilon\,
g(\varepsilon)\,
\im{\left(
\frac{\left[\xi'(p)+p\,\xi''(p)\right]\left[\Delta_{\text{sd}}^2+3(\varepsilon-\xi(p))^2\right]}{4 p[\varepsilon+i 0-\varepsilon^{+}_0(\bb p)]^3[\varepsilon+i 0-\varepsilon^{-}_0(\bb p)]^3}
+2\frac{\left[\xi'(p)\right]^2\left[\Delta_{\text{sd}}^2+(\varepsilon-\xi(p))^2\right] [\varepsilon-\xi(p)]}{[\varepsilon+i 0-\varepsilon^{+}_0(\bb p)]^4[\varepsilon+i 0-\varepsilon^{-}_0(\bb p)]^4}
\right)},
\end{gather}
and the actual value of $W$ is not relevant for the final result. Combining Eqs.~(\ref{Exc_general_0}),~(\ref{calculated_Omega_A_1}), and (\ref{calculated_Omega_A_2}) we find
\be
\label{A1+A2}
\Omega_A[\bb n]=
A_1\left[(\nabla_x\bb n)^2+(\nabla_y\bb n)^2\right]+
A_2\left[\bb n\,\nabla_x^2\bb n+\bb n\,\nabla_y^2\bb n\right]
+W(\bb n\, \nabla_x \bb n)^2+W(\bb n\, \nabla_y \bb n)^2.
\e
Before we proceed, it is important to notice two consequences of the constraint $\bb n^2 \equiv 1$, namely,
\be
\label{wisdom}
\frac{1}{2}\nabla_\beta \bb n^2=\bb n\, \nabla_\beta \bb n=0
\qquad\text{and} \qquad
\frac{1}{2}\nabla_\beta^2 \bb n^2=\nabla_\beta(\bb n\, \nabla_\beta \bb n)=(\nabla_\beta\bb n)^2+\bb n\, \nabla_\beta^2 \bb n=0.
\e
With the help of Eq.~(\ref{wisdom}) we are able to bring Eq.~(\ref{A1+A2}) to the form
\be
\Omega_A[\bb n]=
(A_1-A_2)\left[(\nabla_x\bb n)^2+(\nabla_y\bb n)^2\right],
\e
proving Eq.~(\ref{symmetric_energy_general}) of the main text with $A=A_1-A_2$.

To complete the calculation of the exchange stiffness $A$, one should perform a partial fraction decomposition of the integrands in Eqs.~(\ref{almost_A_1}),~(\ref{almost_A_2}) and make use of the formula
\begin{equation}
\im{\left([\varepsilon-\varepsilon^{\pm}_0(\bb p)+i0]^{-n-1}\right)} =
\frac{(-1)^{n+1}}{n!}\pi\,\delta^{(n)}(\varepsilon-\varepsilon^{\pm}_0(\bb p))
\end{equation}
to integrate over $\varepsilon$ with the result
\begin{multline}
\label{almost_A}
A=\frac{\Delta_{\text{sd}}}{32\pi}T
\int_{0}^{\infty}{d p\,\frac{p\,[\xi'(p)]^2}{\Delta_{\text{sd}}^2}(g_-'-g_+')}
+
\frac{\Delta_{\text{sd}}}{32\pi}T
\int_{0}^{\infty}{d p\,\frac{p\,[\xi'(p)]^2}{\Delta_{\text{sd}}}(g_-''+g_+'')}
\\
+\frac{\Delta_{\text{sd}}}{16\pi}T
\int_{0}^{\infty}{d p\,[\xi'(p)+p\,\xi''(p)](g_-''-g_+'')}
+\frac{\Delta_{\text{sd}}}{16\pi}T
\int_{0}^{\infty}{d p\,p\,[\xi'(p)]^2(g_-'''-g_+''')},
\end{multline}
where $\xi'(p)=\partial \xi/\partial p$ and the derivatives of $g_{\pm}=g(\varepsilon^{\pm}_0(\bb p))=g(\xi(p)\pm\Delta_{\text{sd}})$ are taken with respect to the argument. The latter can also be assumed to be the derivatives with respect to $\xi$,
\be
g_{\pm}^{(n)}=\frac{\partial^{n} g_{\pm}}{\partial \xi^{n}}.
\e
The third term cancels out the fourth term in Eq.~(\ref{almost_A}) after integration by parts with the help of
\be
\xi'(p)+p\,\xi''(p)=\partial [p\,\xi'(p)]/\partial p.
\e
In the remaining terms, one replaces the derivatives of $g_{\pm}=g(\xi(p)\pm\Delta_{\text{sd}})$ with respect to $\xi$ by the derivatives with respect to $\Delta_{\text{sd}}$, reduces the resulting expression to a form of a full derivative with respect to $\Delta_{\text{sd}}$, and uses the relation $\partial g(\varepsilon)/\partial \varepsilon=-f(\varepsilon)/T$ to arrive at Eq.~(\ref{A}) of the main text.





\begin{thebibliography}{99}
\bibitem{comment3}
Strictly speaking, $\bb s$, here, is spin density divided by $\hbar/2$.
\bibitem{comment3.9}
The derivatives of $g(\varepsilon^{\pm}_0(\bb p))$ with respect to the argument were replaced by the derivatives with respect to $\Delta_{\text{sd}}$.
\bibitem{comment4}
We have not yet looked into a rigorous proof of this statement, but perturbative expansion of $\Omega^{\text{DMI}}$ up to higher orders of $\alpha_{\text{\tiny{R}}}$ or with respect to small $J_{\text{sd}}$ together with numerics provide a clear evidence of its validity.
\bibitem{comment5}
DMI in the Bychkov-Rashba model in the presence of linearly polarized light was studied in Ref.~[\onlinecite{Yudin}]. Equation~(10) of that Letter does not show such a cancellation in the absence of electromagnetic field. We argue that it should be corrected.
\end{thebibliography}
\end{document}